\documentclass[letterpaper, 10 pt, conference]{ieeeconf}
\IEEEoverridecommandlockouts
\usepackage{ifpdf}
\usepackage{cite}
\usepackage[hyphens]{url}
\usepackage{stfloats}% this one works for figure* placement
\usepackage{graphicx}
\usepackage[cmex10]{amsmath}
\usepackage{mathtools}% 
\usepackage{algorithmic}
\usepackage[caption=false]{subfig}
\usepackage{array}
\usepackage{color}
\usepackage{amsmath}
\usepackage{soul}

\usepackage{hyperref}
\usepackage[switch]{lineno}
\newcommand{\bi}{\begin{itemize}}
\newcommand{\ei}{\end{itemize}}
\newcommand{\bd}{\begin{displaymath}}
\newcommand{\ed}{\end{displaymath}}
\newcommand{\be}{\begin{equation}}
\newcommand{\ee}{\end{equation}}
\newcommand{\bea}{\begin{eqnarray}}
\newcommand{\eea}{\end{eqnarray}}
\newcommand{\ba}{\begin{array}}
\newcommand{\ea}{\end{array}}
\newcommand{\bc}{\begin{center}}
\newcommand{\ec}{\end{center}}
\newcolumntype{P}[1]{>{\centering\arraybackslash}p{#1}}

\title{\LARGE \bf %The Principle of 
Human Impedance Modulation to Improve Visuo-Haptic Perception}

\author{Xiaoxiao Cheng$^{1,2}$, Shixian Shen$^1$, Ekaterina Ivanova$^{1,3}$, \\ Gerolamo Carboni$^1$, Atsushi Takagi$^4$ and Etienne Burdet$^1$ 
\thanks{$^1$Department of Bioengineering, Imperial College of Science Technology and Medicine, London, UK.}
\thanks{$^2$Department of Electrical and Electronic Engineering, The University of Manchester, Manchester, UK.}
\thanks{$^3$School of Electronic Engineering and Computer Science, Queen Mary University of London, London, UK.}
\thanks{$^4$NTT Communication Science Laboratories, 3‐1 Morinosato Wakamiya, Atsugi, Kanagawa, 243-0198, Japan.} 
\thanks{Emails: \{xiaoxiao.cheng@manchester.ac.uk,  e.burdet@imperial.ac.uk.\} This work was funded in part by the EU H2020 PH-CODING (FETOPEN 829186) and CONBOTS (ICT 871803) grants.}
}

\begin{document}

\maketitle

\begin{abstract}
Humans activate muscles to shape the mechanical interaction with their environment, but can they harness this control mechanism to best sense the environment? We investigated how participants adapt their muscle activation to visual and haptic information when tracking a randomly moving target with a robotic interface. The results exhibit a differentiated effect of these sensory modalities, where participants' muscle cocontraction increases with the haptic noise and decreases with the visual noise, in apparent contradiction to previous results. These results can be explained, and reconciled with previous findings, when considering muscle spring like mechanics, where stiffness increases with cocontraction to regulate motion guidance. Increasing cocontraction to more closely follow the motion plan favors accurate visual over haptic information, while decreasing it avoids injecting visual noise and relies on accurate haptic information. We formulated this active sensing mechanism as the optimization of visuo-haptic information and effort. This OIE model can explain the adaptation of muscle activity to unimodal and multimodal sensory information when interacting with fixed or dynamic environments, or with another human, and can be used to optimize human-robot interaction.
\end{abstract}
\noindent \textit{Author summary}: \\
It has been widely known that the human CNS stiffens the limbs in response to movement error due to interactions. However, by systematically investigating how muscle co-contraction varies with incoming visual and haptic information, we show that this principle does not always hold, and limb stiffness is in fact regulated by the brain to optimise incoming sensory information from the interaction. The  manuscript thus describes the following contributions:
\begin{itemize}
    \item Muscle cocontraction (from which the limb stiffness depends) changes in accordance with the quality of visual and haptic cues. It relaxes with increasing visual noise and stiffens with haptic perturbation, contrary to suggestions from previous studies.
    \item We propose a mathematical model of this active sensing mechanism, according to which muscle co-contraction is adapted by the brain to minimise prediction error about the environment based on visual and haptic information, while saving effort.
    \item This computational model: i) predicts the observed results both qualitatively and quantitatively; ii) can explain the results on cocontraction adaptation to force field and interaction with human partners that we know from the literature; iii) extends the existing computational models, and can be used to improve human-robot interaction.
\end{itemize}

\section{Introduction}
How do humans interact with their environment? It is known that the central nervous system (CNS) regulates the limbs’ stiffness by coordinating muscle activation to shape the energy exchange with the environment \cite{Hogan1984, Burdet2001}, {\color{black} such as unstable situations typical of tool use \cite{Franklin2007, Selen2009}. The prevailing explanation for the observed adaptation of muscle cocontraction is that individuals adjust stiffness to minimize errors in the presence of disturbances \cite{Franklin2003, tee2010concurrent, yang2011human}.} However, how this affects visuo-haptic sensing has been little investigated. For instance, when skiing down a bumpy slope, should one stiffen the legs to best sense the terrain, or relax them to filter out perturbations? This may be particularly important when visual information is degraded, such as in foggy conditions or at dawn, where one must rely on one's feet to feel the terrain and avoid falling.

{\color{black} Few studies have examined how muscle stiffness regulation is influenced by visual disturbances, and the results have shown complex response patterns. For example, \cite{osu2004optimal} found that in arm reaching tasks, endpoint stiffness decreased with larger target sizes, indicating that the CNS increases stiffness to enhance control precision. 
However, endpoint stiffness did not significantly increase in response to lateral visual noise during arm reaching tasks, unlike the increase observed with mechanical vibrations \cite{wong2009influence}. This suggests that the CNS may respond differently to visual versus haptic disturbances. Further research is needed to explore how visual disturbances affect motion control through muscle stiffness regulation.}

{\color{black} Visual and haptic information are critical in physical human-robot collaboration, including applications such as physical rehabilitation \cite{colombo2018rehabilitation}, collaborative robotics for industrial manufacturing \cite{colgate2003intelligent}, and shared control of semi-autonomous vehicles \cite{mulder2012sharing}. However, how the CNS combines these sensory inputs in real time remains unclear. 
When integrating sensory signals over short intervals, the CNS accounts for both sensory discrepancies and temporal delays to achieve optimal multi-sensory integration and feedback control \cite{crevecoeur2016dynamic}. Interestingly, the presence of visual feedback during a mechanical disturbance does not increase the magnitude of the muscle response but does reduce its variance \cite{kasuga2022integration}. Additionally, we observed a modulation of coactivation when physically connected individuals track a common target \cite{Hendrik2022}. The partner with superior visual acuity tends to stiffen their arm and lead the movement, while the other relaxes their arm. Notably, the partners adjust their cocontraction differently depending on the levels of visual and haptic noise.}
%figure
\begin{figure*}[!t]
\centering
\includegraphics[width=1\textwidth]{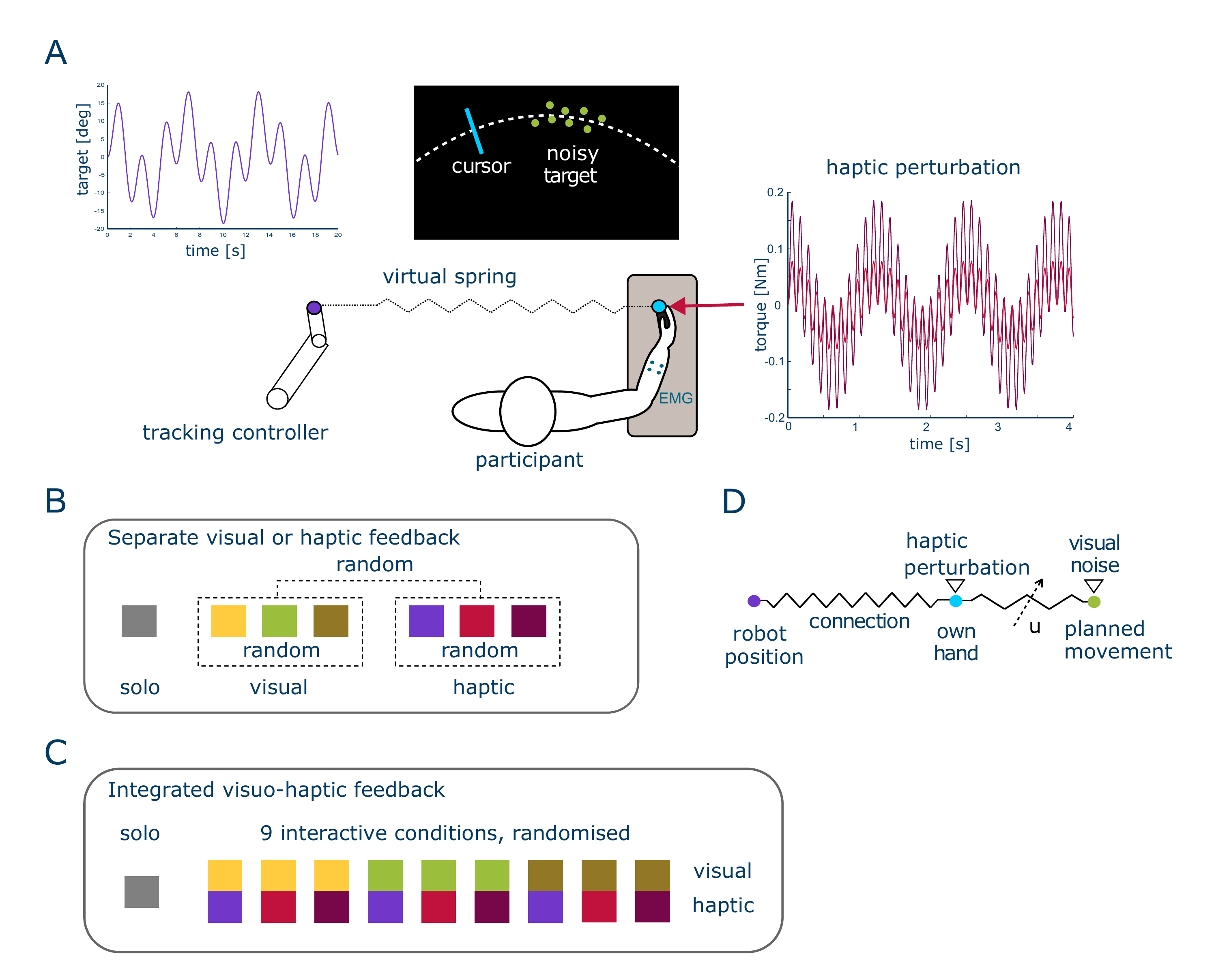}
\caption{\label{f:experiment}
Experiment setup and protocol. A: Participants were asked to track a randomly moving target with noisy visual feedback and in some conditions were connected to the human-like tracking controller of \cite{Takagi2017}. B,C: Experiment protocol of separate visual or haptic feedback experiment, with each block consisting of nine trials. The 13 participants received only visual/haptic feedback in random order and each with random noise level (B). Another 22 participants experienced nine integrated visual and haptic conditions presented in a random order (C). (D) illustrates the mechanical modelling scheme of the human-robot interaction with visual and haptic noise.}
\end{figure*}

In order to systematically study how humans adapt their muscle activation with visual and haptic feedback, we conducted an experiment in which subjects tracked a randomly moving target using wrist flexion/extension while being connected to the human like tracking controller of \cite{Takagi2017} (Fig.\,\ref{f:experiment}A). We examined the influence of visual and haptic feedback with different levels of noise first separately (Fig.\,\ref{f:experiment}B), then in combination (Fig.\,\ref{f:experiment}C). In the visual conditions the target presented on the monitor was either a sharp disk, or a dynamic cloud of normally distributed dots. We also introduced a haptic perturbation of varying amplitude to the interaction torque. Conditions with a specific noise level were presented pseudo-randomly. We observed that the visual and haptic noise levels have different effects on the cocontraction adaptation, suggesting that the brain modulates body impedance based not only on movement error \cite{Franklin2008}, but also on its influence on specific sensory interactions. Subsequently, we developed a computational model to examine the mechanism behind muscle cocontraction adaptation (Fig.\,\ref{f:experiment}D).
%figure
\begin{figure*}[!t]
\centering
\includegraphics[width=\textwidth]{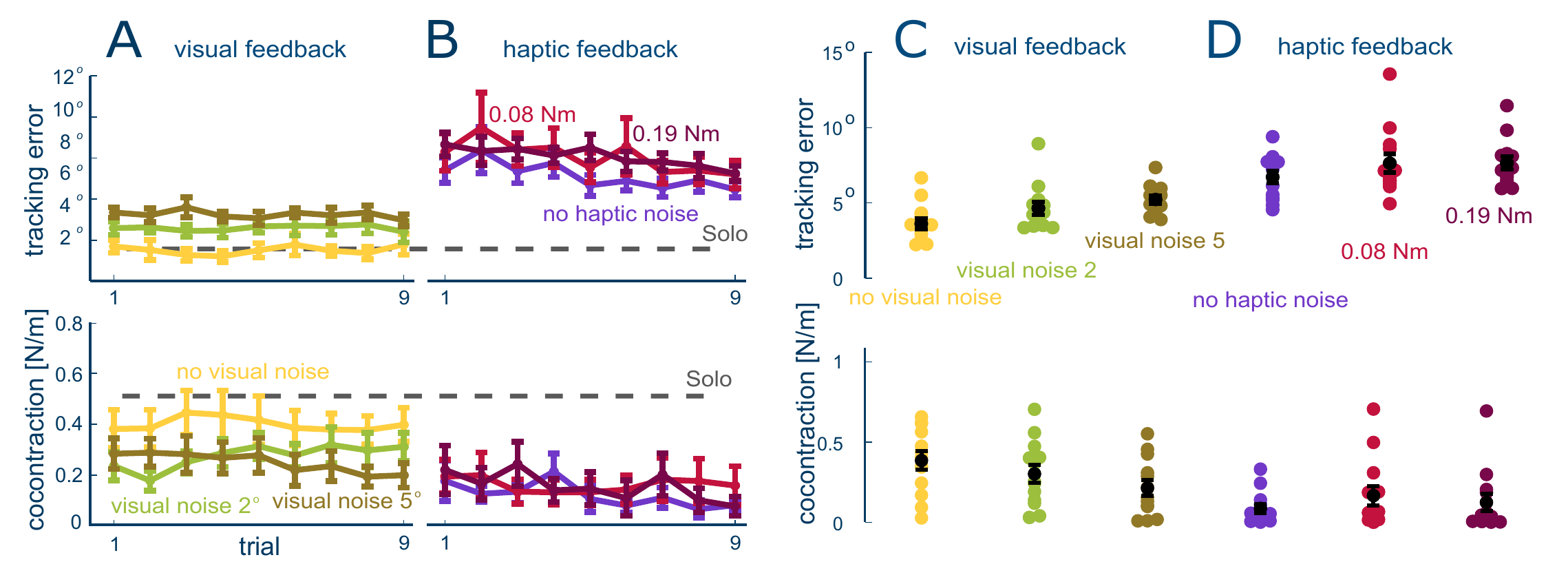}
\caption{\label{f:results_sole}
Results of solely visual/haptic feedback experiment. A\&B: Evolution of tracking error and cocontraction with visual and/or haptic feedback, where error bars represent standard error. {\color{black} C\&D: The mean and standard error of tracking error and cocontraction for all subjects during the last four trials.} In the visual feedback condition (A\&C), with increasing visual noise the tracking error increases while the cocontraction decreases. In the haptic feedback condition (B\&D), there is a decaying trend of both tracking error and cocontraction over the trials and no clear difference among noise conditions. 
}
\end{figure*}

%figure
\begin{figure*}[!t]
{\color{black}
\centering
\includegraphics[width=\textwidth]{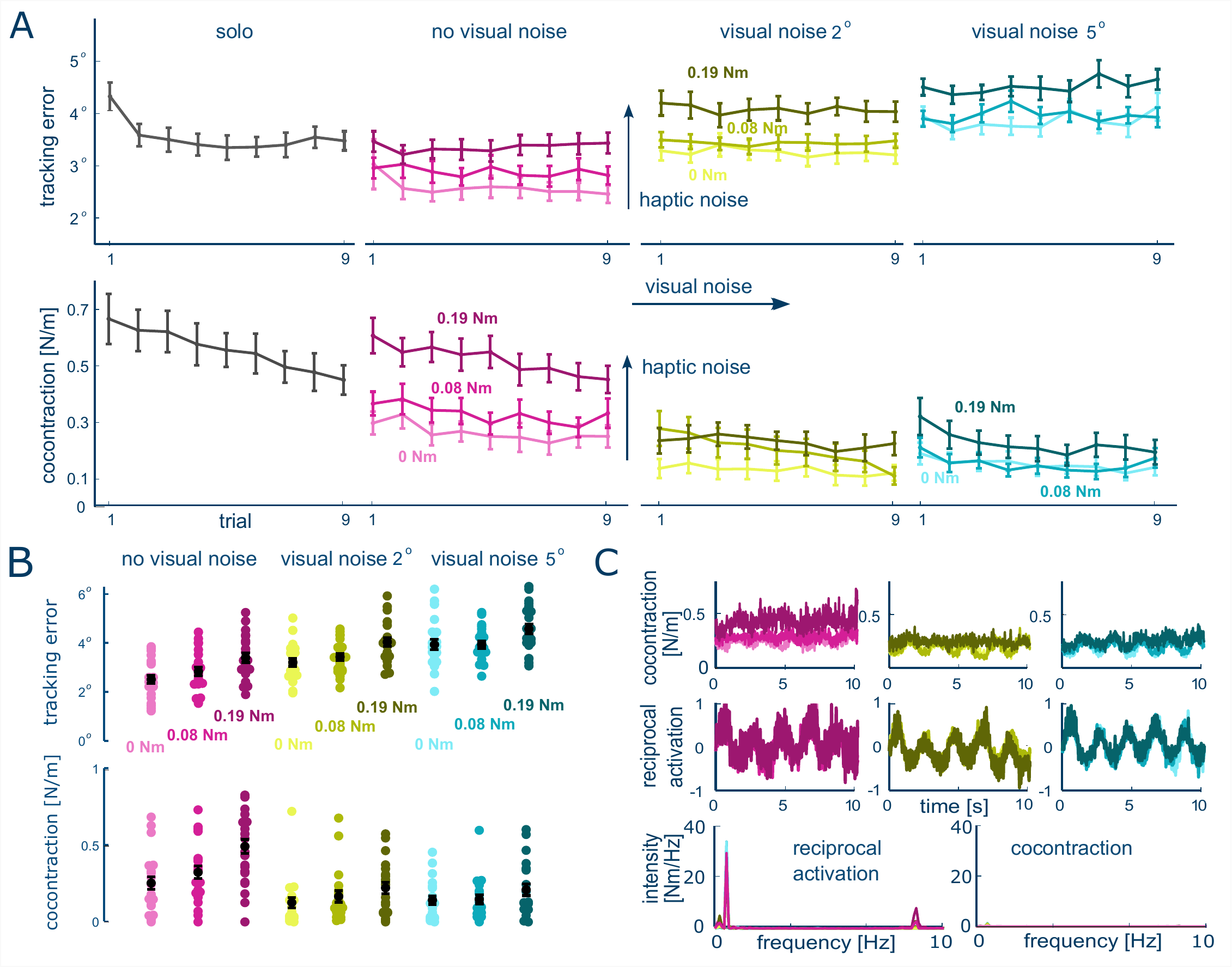}
\caption{\label{f:results}
Results of tracking with combined visual and haptic noise. A: Evolution of tracking error and cocontraction with visual and/or haptic feedback, with error bars representing the standard error. The tracking error saturates in the initial solo trials, and increases with both visual and haptic noise during interactive trials. Cocontraction shows a slower decrease across trials. 
B: Mean and standard error of tracking error and cocontraction for all subjects during the last four trials. Tracking error increases with either visual or haptic noise, while muscle cocontraction increases with haptic noise and decreases with visual noise. C: Muscle cocontraction and reciprocal activation waveforms along with their frequency spectrum. Muscle cocontraction remains relatively constant, while reciprocal activation changes synchronised with the movement, as is confirmed by their spectra.}
}
\end{figure*}

%%%%%%%%%%%%%%%%%%%%%%%%%%%%%%%%%%%%%%%%%%%%
\section{Results}
We first analyzed how visual feedback affects the motion control (Fig.\,\ref{f:results_sole}). After the initial solo trials with only visual feedback, the tracking error remains stable in each of the three visual feedback conditions (slope $|s|<0.03,\, p>0.50$). However, the tracking error increases with the magnitude of visual noise (larger error in each of the weak and strong conditions relative to clean vision, $p<0.01$, {\color{black} pairwise} Wilcoxon tests). On the other hand, the cocontraction level decreases with larger visual noise ({\color{black}$p<0.001$ for strong noisy condition relative to sharp vision, $p = 0.04$ for weak noisy condition compared to sharp vision, paired t-test} ).

These results seem to contradict previous observations that muscle cocontraction increases with the magnitude of movement error  \cite{Franklin2004, wong2009influence, franklin2012visuomotor, calalo2023sensorimotor}. However these earlier findings can be explained when considering the spring like muscle mechanics \cite{burdet2013human}, where muscle activation increases stiffness and viscosity while also shortening the muscle. By coordinating the activation of antagonist muscles, the CNS can thus control the force at the hand as well as the spring stiffness and reference position. 
{\color{black} In particular we can consider the reference position that emerges when the CNS controls muscles' activity to move the limbs. 
As illustrated in Fig. \ref{f:experiment}D, with clear visual information, increasing muscle coactivation will increase the limb's endpoint stiffness and guide it closer to this accurate \textit{motion plan}. However, when the target information becomes noisy, stiffening the muscle would instead inject noise into the limb’s movement. This explains why cocontraction decreases with an increasing level of noise in the visual target.}

Next we investigated how the CNS regulates muscle cocontraction when both visual and haptic feedbacks are provided. Fig.\,\ref{f:results}A shows that the error decreases fast in the solo trials (slope $s=-0.06,\,p=0.0002$) and reaches a steady level ($s=0.01,\,p=0.64$ for the last seven trials). The level of error remains stable in all the interactive conditions (non-significant slope $s \neq 0$ with $p>0.05$). A two-way ART ANOVA shows that the tracking error depends on both visual noise ($F(2,38)=85.824, p<0.0001$) and haptic noise ($F(2,38)=80.673, p<0.0001$), increasing with the amplitude of visual or haptic noise ({\color{black} }$p<0.001$ {\color{black}for all pairwise comparisons between noise levels,} Wilcoxon tests).

The muscle cocontraction tends to decrease with trials, indicating a learning effect to integrate the two sensory feedback modalities (especially at the largest level of haptic noise with $s=-0.0226,\,p<0.0001$ for sharp vision). Consistent with the sole visual feedback condition in the previous experiment, cocontraction decreases with the visual noise as shown in Fig.\,\ref{f:results}C {\color{black}(significant interaction between visual and haptic noise factors $F(4,76)=5.1436, p=0.001$)}, where the sharp vision conditions results in a higher cocontraction level compared to both weak and strong visual noise conditions {\color{black} for each level of haptic noise} (all $p<0.003$, {\color{black}} Wilcoxon tests). However, muscle cocontraction increases with the level of haptic perturbation, especially between weak and strong haptic perturbation conditions. The increment of this increase depends on the visual noise level: it increases the most in the sharp visual condition ({\color{black}$p<0.05$, Wilcoxon tests}) and becomes less clearly with the increase of visual noise ({\color{black} all $p>0.05$, Wilcoxon tests}). 

In order to understand how the activity of antagonistic muscle is modulated dynamically, we align the reciprocal activation and cocontraction profiles as shown in Fig.\,\ref{f:results}C. Muscle cocontraction remains relatively constant for each sensory noise condition (slope $|s|<0.008$ for all conditions). On the other hand, the reciprocal activation is modulated to produce movement dynamics (the averaged Pearson correlation coefficient between reciprocal activation and target movement is {\color{black} 0.82$\pm$0.09}). The frequency spectrum of reciprocal activation has {\color{black} three} peaks at the target movement frequencies (0.2, 0.5 {\color{black} and 8.5\,Hz}) in contrast to the essentially flat spectrum of muscle cocontraction. 
These results indicate that reciprocal activation generates the tracking movement {\color{black} and responds to the haptic perturbation}, while the cocontraction level is regulated to deal with the specific noise condition.

%%%%%%%%%%%%%%%%%%%%%%%%%%%%%%%
\section{Modelling of visuo-haptic sensing}
What are the principles of the cocontraction adaptation? Above experiments show that cocontraction tends to decrease with practice, and is modulated by both visual noise of the target, and external haptic noise. However, these two noise sources have an opposite effect: Cocontraction increases with a larger level of visual noise, but decreases with haptic noise. We posit that these apparently contradictory trends can be explained through the following sensorimotor interaction principles:
\begin{itemize}
\item Muscles' activation generation corresponds to the CNS using reciprocal activation to move the limbs and coactivation to guide them along a motion plan with suitable viscoelasticity \cite{Hogan1984}.
\item Cocontraction is adapted to maximise performance considering both visual target noise and haptic noise at the limbs while concurrently minimising effort.
\end{itemize}
The mechanics of these principles can be illustrated as in Fig.\,\ref{f:experiment}D, where muscle cocontraction can tune the viscoelasticity of the hand to follow the planned movement. When the target is visually sharp, the motion plan is accurate thus it is useful to stiffen in order to follow it closely. However when the target is noisy, it is preferable to relax in order to avoid injecting own visual noise into the hand movement and benefit from the external haptic guidance.

These principles can be formulated mathematically by considering the maximal likelihood \textit{prediction error} when integrating visual and haptic information:
\be
\Gamma(u) \equiv \, \frac{\sigma_t^2(u) \,\, \sigma_h^2}{\sigma_t^2(u) + \sigma_h^2}
\label{e:predictionError}
\ee
with $\sigma_h$ the standard deviation of the haptic noise exerted on the limb. Critically, the standard deviation $\sigma_t(u)$ of hand movement relative to the motion plan following the target can be regulated through the muscle co-contraction $u$.

The above sensorimotor interaction principles thus correspond to the concurrent minimization of the prediction error and effort $u^2$, i.e. of the cost function
\be \label{eq:cost}
V(u) = \Gamma(u) + \frac{\gamma}{2} u^2\,, \quad \gamma>0 
\ee
with the effort ratio $\gamma$. Muscle cocontraction
can then be adapted using a gradient descent optimisation:
\bea
u_{new} \!\!\!&=&\!\!\! u - \frac{dV(u)}{du} \,= \,- \frac{d\Gamma(u)}{du} \,+\, (1-\gamma)\,u \,, \nonumber \\
-\frac{d\Gamma(u)}{du} \!\!\!& = &\!\!\! \left[ \frac{\sigma_h^2}{\sigma_t^2 (u) +\sigma_h^2}\right]^2 \left[-\frac{d \sigma_t^2(u)}{du} \right] >0 \,.
\eea 
%figure
\begin{figure*}[!t]
{\color{black}
\centering
\includegraphics[width=1\textwidth]{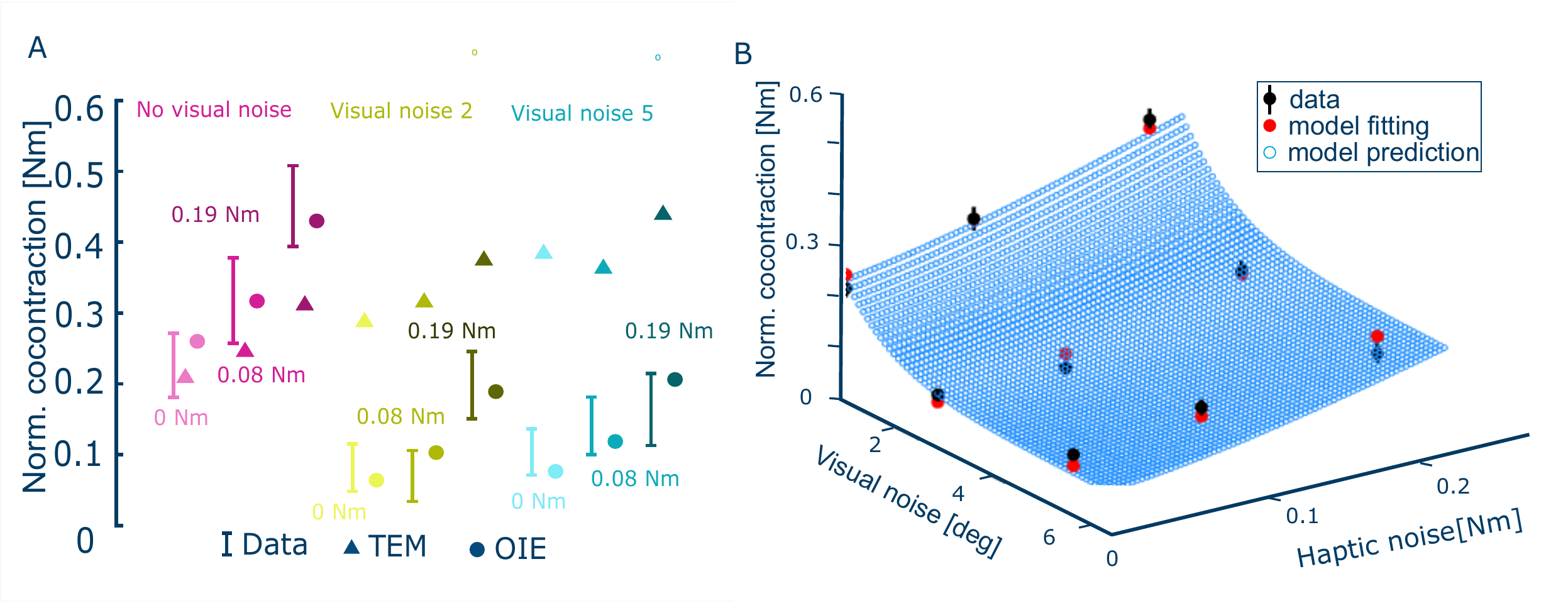}
\caption{\label{f:surface}
Simulation results of normalized cocontraction. (A) Comparison between the \textit{tracking error minimization} (TEM) model and our \textit{optimal information and effort} (OIE) model across the nine experiment conditions. The OIE model predicted normalized cocontraction values closely aligned with the experimental data, while the TEM model produced large prediction errors. (B) OIE model predictions of normalized cocontraction as a function of visual and haptic noise levels. Black dots represent the recorded average cocontraction of 20 participants in the final trial. Red dots represent the fitted data from the OIE model, and green dots the predicted data. The blue dots show the model's predicted cocontraction in unobserved noise conditions.}
}
\end{figure*}

This \textit{optimal information and error} (OIE) model was tested on the data of the tracking experiment with combined visual and haptic noise. The effective noise values that best fit the nine data points of Fig.\,\ref{f:surface} are given in Table\,\ref{Table:noise}, yielding an effort ratio {\color{black}$\gamma^*$\,=\,2.26}. The model predicted muscle cocontraction values are shown in Fig.\,\ref{f:surface}. {\color{black} We also tested the \textit{tracking error minimisation} (TEM) model of \cite{Franklin2008} that explains the motor learning in novel force fields. In this model, the coactivation $u$ increases with each new trial to minimize tracking error $e$, and decreases to minimize effort, according to
\be \label{eq_Vshape}
u_{new} \equiv \, \alpha \, e + (1-\gamma)\,u \,, \quad 0<\alpha\,, \quad 0<\gamma <1 \, .
\ee

The predicted values from the OIE and TEM models are compared with experimental data as shown in Fig.\,\ref{f:surface}A. The results indicate that the TEM model correctly predicts the trend of increased cocontraction with haptic noise, it fails to capture the decrease in cocontraction with visual noise. In contrast, the OIE model closely matches the experimentally measured cocontraction levels in all variations of visual and haptic noise. 
The superiority of the OIE model is further supported by the Akaike information criterion (AIC): the small sample-size normalised AIC value \cite{portet2020primer, cohen2021normalized} for the OIE model is -6.5, which is lower than the value of -2.9 for the TEM model, showing that the OIE model better accounts for information loss and the number of independent parameters. Fig.\,\ref{f:surface}B illustrates how the OIE model accurately predicts the specific modulation of cocontraction to varying levels of visual and haptic noise.}
%table
\begin{table}[t]
  \centering
  \caption{Identified effective visual and haptic noise values}
  \label{Table:noise}
\begin{tabular}{|P{4cm}|| P{0.8 cm}|P{0.8 cm}|P{0.8 cm}|}
%  \hline
%  \multicolumn{4}{|c|}{Comparison between model predicted values and experiment data of muscle cocontraction} \\
 \hline
 Noise strength & sharp & weak & strong\\
 \hline
{\color{black} Effective haptic perturbation $\sigma_h*$} & {\color{black}5.06} & {\color{black}5.86} & {\color{black}7.85} \\
 {\color{black}Effective visual noise $\sigma_v*$} & {\color{black}30.64} & {\color{black}63.66} & {\color{black}65.30} \\
%  Prediction error [Nm] & 0.0177  &  0.0087 &   0.0171 &   0.0055 &    0.0188 &    0.0262 &    0.0202 &    0.0047 &    0.0312\\
 \hline
\end{tabular}
\end{table}

%table
% \begin{table*}[ht!]
%   \centering
%   \caption{Comparison of muscle cocontraction (in [Nm]) between model predicted values and experimental data }
%   \label{Table:comparison}
% \begin{tabular}{ |P{2.8cm}||P{0.7cm}|P{0.7cm}|P{0.7cm}|P{0.7cm}|P{0.7cm}|P{0.7cm}|P{0.7cm}|P{0.7cm}|P{0.7cm}|  }
%  \hline
%  Visual \& haptic noise & V0H0 & V1H0 & V2H0 & V0H1 & V1H1 & V2H1 & V0H2 & V1H2 & V2H2 \\
%  \hline
%  Experimental data &   0.24   & 0.10 &   0.10 &   0.29 &   0.10 &    0.14 &    0.46 &    0.25 &    0.19
% \\
%  Model predicted values &     0.24 &   0.11  &  0.10 &   0.28 &   0.14  &  0.13  &  0.43 &   0.25 &    0.25
% \\
% \hline
% \end{tabular}
% \end{table*}

Could the OIE model predict the effect of blind haptic interaction? In this case the lack of visual feedback can be modeled through visual noise with infinite deviation $\sigma_t \rightarrow \infty$, thereby making the cost to minimize $V(u) = \sigma_h^2+u^2$. The OIE model then predicts that subjects connected haptically to the controller tracking the target (but without visual feedback) would minimize cocontraction independent of the haptic noise level. The results of the experiment testing this prediction are shown in Fig.\,\ref{f:results_sole}B. In the sole haptic feedback conditions, both tracking error and muscle cocontraction decreased over the trials ($F(1,12) = 10.70,, p<0.01$, two-way ART ANOVA comparing the mean of all subjects in the first and last three trials). Consistent with the model prediction, muscle cocontraction remained at a minimum value and did not depend on the noise level ($p>0.30$, paired Wilcoxon tests). There was little change in tracking error with the haptic perturbation level ($F(2,24) = 1.30, , p=0.28$, two-way ART ANOVA).

%%%%%%%%%%%%%%%%%%%%%%%%%%%%%%
\section{Discussion}
{\color{black}In physical human-robot interaction, sensory signals crucial for motor control are derived from both visual input and haptic feedback, which provides substantial information about the movement intentions of the connected agents. However, the process by which the CNS adjusts muscle cocontraction to optimize the use of visual and haptic signals, thereby enhancing the effectiveness of human-robot collaboration, remains unclear. This paper systematically investigated how different levels of visual and haptic disturbances affect muscle activation during target-tracking tasks.

In a first experiment, we evaluated the impact of noise on muscle control within either the visual or haptic channel. Contrary to previous observations\cite{Franklin2008}, muscle co-activation decreased with increasing levels of visual noise. Meanwhile, muscle co-activation showed no substantial variation in response to the intensity of haptic disturbances. This suggests that the CNS regulates muscle co-activation by considering not just movement error but also sensing uncertainty. When there is no visual noise and motion planning can be relied upon, muscle cocontraction increases to ensure that the planned trajectories are well followed. Conversely, as visual noise and the associated uncertainties in motion planning increase, muscle co-activation decreases. 
In the absence of visual feedback, which can be interpreted as maximal uncertainty from visual sensing, muscle cocontraction remains low and becomes insensitive to the intensity of haptic disturbance.

The second experiment extended these findings by examining the influence of noise in scenarios with both visual and haptic feedback. Consistent with the observations in the single feedback conditions, muscle cocontraction decreased with an increase in visual noise and increased with rising haptic disturbances, in line with previous studies \cite{Franklin2004, tee2010concurrent}. 
This indicates that the cocontraction adaptation mechanism is influenced by both visual and haptic feedback. When visual feedback is clear, muscle cocontraction significantly increases with rising haptic noise; however, this response diminishes markedly when visual feedback is blurred. 
This behavior suggests that the CNS's reliance on a particular sensory modality correlates with the uncertainty associated with each feedback type, revealing a complex, nonlinear interplay between these modalities. Notably, muscle co-activation drops sharply at the onset of visual noise but decreases as visual noise intensifies. Additionally, muscle co-activation shows a general downward trend with an increasing numbers of experimental trials, likely reflecting the CNS’s strategy to minimize metabolic cost \cite{Todorov2002, Franklin2008}.}

By systematically studying how subjects interact with visual and haptic information, we were able to decipher the mechanism of body impedance adaptation during the interaction with the environment. Our experimental and simulation results demonstrated that subjects regulate co-activation to optimally integrate visual and haptic information while minimizing effort. 
This enables them to extract useful information about their environment and plan accurate movements. A computational model based on the optimization of information and effort (OIE) was used to predict muscle cocontraction levels under different visuo-haptic sensory conditions. This OIE model explains how the CNS integrates multi-modal sensory information, considering their respective noise level, to enhance perceptual acuity with minimal metabolic cost. Notably, the results obtained could not be predicted by previous models of muscle cocontraction adaptation \cite{Franklin2008, Theodorou2010, Li2018a}, which only considered movement error and suggested that muscle activation would increase with either visual or haptic noise. In contrast, the OIE model accounts for sensorimotor interactions and the influence of each sensory modality's noise on target perception, tracking performance, and effort, thereby successfully predicting the observed actions.

{\color{black} 
The OIE model can also explain the experimental results presented in \cite{osu2004optimal}, where participants marginally reduced stiffness in response to a larger target size, resulting in increased visual uncertainty.} It is also compatible with existing models for motor adaptation to external haptic \cite{Burdet2001, Franklin2004, franklin2007endpoint} or visual \cite{wong2009influence, franklin2012visuomotor, calalo2023sensorimotor} perturbations. A destabilizing force field \cite{Burdet2001, franklin2007endpoint}, or perturbations of the hand position in visual feedback \cite{wong2009influence, franklin2012visuomotor, calalo2023sensorimotor}, both correspond to haptic noise and result in increased stiffness, as predicted by the OIE model. In turn, this means that the OIE model extends the computational models of \cite{Franklin2008, Theodorou2010} according to which stiffness increases with hand movement error.

In conclusion, muscle cocontraction is not automatically tuned to minimize movement error as previously thought \cite{Franklin2008, Theodorou2010, Li2018a}. Instead, it is skillfully regulated by the CNS to extract maximal information from the environment. The OIE model presented in this paper explains the adaptation of muscle cocontraction during interactions with force fields, dynamic environments with visual noise at the target, and haptic noise at the hand, as in the experiments of this paper, as well as during collaboration with other humans \cite{Hendrik2022}. {\color{black}While active sensing has been identified previously in vision \cite{Rao1999}, this is, to our knowledge, the first evidence of body adaptation to improve visuo-haptic sensing. Furthermore, while sophisticated algorithms have been developed for such active inference e.g. \cite{Friston2010}, the experimental results were well predicted through the simple OIE model.} This active sensing computational mechanism may be used to design robotic systems that adapt to their user as a human partner would, with various applications in physical human-robot collaboration.

%%%%%%%%%%%%%%%%%%%%%%%%%%%%%%%%%%%%%%
\section{Materials and Methods}   \label{s:experiment}

\subsection{Experiment setup}
Each participant was seated on a height-adjustable chair, next to the Hi5 robotic interface \cite{Melendez-Calderon2011} with the dominant wrist attached to a handle of the interface during flexion/extension movement. They received visual feedback of the target angle and of their wrist flexion/extension angle on their monitor, and/or haptic feedback from the interaction with the tracking controller of \cite{Takagi2017} (Fig.\,\ref{f:experiment}A).

The Hi5 handle is connected to a current-controlled DC motor (MSS8, Mavilor) that can generate torques of up to 15\,Nm, and is equipped with a differential encoder (RI 58-O, Hengstler) to measure the wrist angle and a torque sensor (TRT-100, Transducer Technologies) to measure the exerted torque in the range [0,11.29]\,Nm. The handle is controlled at 1\,kHz using Labview Real-Time v14.0 (National Instruments) and a data acquisition board (DAQ-PCI-6221, National Instruments) while the data was recorded at 100\,Hz.

The activation of two antagonist wrist muscles, the flexor carpi radialis (FCR) and extensor carpi radialis longus (ECRL) were recorded during the movement from each participant. Muscle electromyographic (EMG) signals were measured with surface electrodes using a medically certified non-invasive 16-channel EMG system. The EMG data was recorded at 100\,Hz. The raw EMG signal was \textit{i}) high-pass filtered at 20\,Hz by using a second-order Butterworth filter to remove drifts in the EMG and \textit{ii}) rectified and passed through a low-pass second-order Butterworth filter with a 15\,Hz cutoff frequency to obtain the envelope of the EMG activity.

\subsection{Tracking task} \label{section:tracking}
{\color{black}The participants were instructed to ``track the moving target as accurately as possible'' using wrist flexion-extension movements.} The target was moving according to 
\begin{eqnarray}
q^*(t) \!\!\!&\equiv&\!\!\!  \, 18.5 \, \sin \! \left(2.031\,t^* \!\right)  \, \sin\!\left(1.093\,t^*\!\right) \nonumber \\
t^* \!\!\!&\equiv&\!\!\! t + t_0 \,\,, \quad 0\leq t\leq 20\,s
\label{eq:b_traj}
\end{eqnarray}
where $t^*$ started in each trial from a randomly selected offset time $\{t_{0} \in [0, 20]\,\mathrm{s}\ |\ q^*(t_{0}) \equiv 0\}$ of the multi-sine function in order to minimize memorization of the target's motion.

In \textit{solo trials}, the participants tracked the target without active torque from the robot. Otherwise, participants’ wrist was connected by a compliant virtual spring to the tracking controller of \cite{Takagi2017} according to (in Nm)
\be
\tau(t) = \,0.03 \,[q_c(t)-q(t)] \, ,
\ee
where $q$ (in degrees) denotes the participant’s wrist angle, and $q_c$ the controller's target angle computed as in \cite{Takagi2017}. The connection stiffness was selected such that subjects could clearly sense the robot's movement but compliant enough to let them actively pursue the tracking task \cite{ivanova2020motion}. {\color{black} This controller has been shown to induce a similar behaviour to human interaction \cite{ivanova2020motion}. Using this human-like interaction (rather than direct human interaction) allows for the direct manipulation of haptic noise. }

The experiment considered three haptic noise conditions: \textit{sharp haptic information} (H0 condition) without noise, or a torque perturbation $\sigma_p \sin(25\,t) \sin(30\,t),\, 0\leq t \leq 20s$
with $\sigma_p = 0.08$\,Nm in the {\it{weak haptic noise condition}} H1 and $\sigma_p = 0.19$\,Nm in the {\it{strong haptic noise condition}} H2. 

For visual feedback, a target was displayed on the monitor for participants to track with three conditions. In the {\it{sharp visual condition}} V0, the target was displayed as a 8\,mm diameter disk, which had the same visual condition as in solo trials. In the {\it{noisy visual conditions}} V1 and V2, the target was displayed as a ``cloud’’ of eight randomly distributed dots around the nominal target position, where each of the eight dots was sequentially replaced every 100\,ms. The cloud's vertical position relative to the target was normally distributed with {\color{black}$\eta \in \mathcal{N}$[0,\,(15\,mm)$^2$]}, the angular distance position relative to the target was distributed with $\eta_q \in \mathcal{N}$[0,\,$\sigma_c^2$], and the velocity with {\color{black}$\eta_{\dot{q}} \in \mathcal{N}$[0,\,(101.6\,mm/s)$^2$]}. The {\it{weak visual noise condition}} V1 was defined through {\color{black}$\sigma_c\,$=\,21.32\,mm} and the {\it{strong visual noise condition}} V2 through {\color{black}$\sigma_c$ = 52.78\,mm.} 
% 1 deg = 10.15 mm

\subsection{Muscle activation calibration} \label{subsec:muscle}
The participants placed their wrist in the most comfort- able middle posture which defined 0$^\circ$. Constrained at that posture, they were then instructed to sequentially flex or extend the wrist to exert torque. Each phase was 4\,s long and was followed by a 5\,s rest period to avoid fatigue. The latter period was used as a reference activity in the relaxed condition. This procedure was repeated four times at flexion/ extension torque levels of \{1,\,2,\,3,\,4\}\,Nm and \{-1,\,-2,\,-3,\,-4\}\,Nm, respectively.

The recorded muscle activity of each participant was then linearly regressed against the torque values. Specifically, the torque of the flexor muscle was modelled from the envelope of the EMG activity $u_f$ as
\begin{equation}
	\begin{aligned}
		\tau_f(t) = \, \alpha_0 \, u_f(t) \, + \, \alpha_1 \,, \quad \alpha_0, \alpha_1 > 0 \,,
	\end{aligned}
	\label{eq:regression}
\end{equation}
and similarly for the torque of the extensor muscle $\tau_e (t)$.

\subsection{Experimental protocol}
The experiment was approved by the Imperial College Research ethics committee (No. 15IC2470). %and eight participants (two females, 20-26 years old) carried out the additional frequency experiment. 
Before participating in the experiment, each subject was informed about the investigation, then signed a consent form, filled out a demographic questionnaire and an Edinburgh Handedness Inventory form \cite{Oldfield1971}.

After the EMG calibration, the participants carried out nine solo trials to get familiar with the tracking task and the dynamics of the wrist interface. This was followed by series of 20\,s long trials as described in Fig.\,\ref{f:experiment}B. After each trial, the target disappeared, and the participants had to place their respective cursor on the starting position at the center of the screen. The next trial then started after a 5\,s rest period and a 3\,s countdown. The participants could take an extra break at will between trials by keeping the cursor away from the screen center.

\subsubsection{Behaviour with only visual or only haptic feedback}
13 naive subjects (six females and seven males) aged 21-25 years (mean = 22.5, sd = 1.05) were recruited to study the influence of visual or haptic feedback separately. One participant was left-handed (with Laterality Quotient LQ $=$ -43) and the others were right-handed (LQ $>$ 40). 

Each participant carried out two blocks with visual feedback, respectively haptic feedback, presented in a random order. Furthermore, the noise conditions were presented randomly in each block. Nine interaction trials of 20 seconds were carried out in each of these conditions. 

\subsubsection{Behaviour with visual and haptic feedback}
For the coupled visuo-haptic feedback experiment, 22 naive subjects (twelve females and ten males) aged 22-35 years (mean = 24.1, sd = 3.06) were recruited to study the combined effect of visual and haptic feedback. One participant was ambidextrous (LQ\,=\,-29) and the others were right-handed (LQ\,$>$\,40). {\color{black} Due to incomplete EMG data for two participants, data analysis was conducted using the remaining 20 subjects.}

There were nine blocks of nine interaction trials, each with one of the nine different noise conditions (resulting from the combination of visual the three noise levels and three haptic perturbations) presented in a random order. 

%%%%%%%%%%
\subsection{Analysis}
The tracking error and muscle cocontraction were used as metrics to analyze the participants’ tracking performance and impedance adaptation to different visuo-haptic noise conditions. To represent the overall tracking accuracy within a trial, the \textit{tracking error} is defined as the root mean squared error between the target position $q^* (t)$ and the hand position $q(t)$ during one trial:
\be
e \equiv \, \left(\!\frac{1}{T} \! \int_0^T \!\!\! \left[q^*\!(t) - q(t) \right]^2 dt \!\right)^\frac{1}{2} \!\!,\, \,\,\, T \equiv 20\,s.
\label{e:error}
\ee
Using the torque regression model of eq.\,(\ref{eq:regression}), the joint \textit{reciprocal activation} is defined as
\begin{equation}
\tau(t)= \tau_{\rm f}(t) - \tau_{\rm e}(t)
\end{equation}
and \textit{cocontraction} as
\begin{equation}
u(t)= \, \min\{\tau_{\rm f} (t),\tau_{\rm e} (t)\}\,.
\end{equation}
The average muscle cocontraction $\Bar{u}$ of each trial for a specific subject is computed as
\begin{equation}
\Bar{u} = \, \frac{1}{T} \!\! \int_0^T \!\!\!\! u(t) \, dt\,, \quad T = 20\,s.
\end{equation}
The \textit{normalised cocontraction} of each participant was used in the subsequent analysis, which is calculated as
\begin{equation}
    u_{n} = \frac{\Bar{u}-\Bar{u}_{min}}{\Bar{u}_{max}-\Bar{u}_{min}}\,,
\end{equation}
where $\Bar{u}_{min}$ and $\Bar{u}_{max}$ are the minimum and the maximum of the averaged muscle cocontraction of all interaction trials of a participant respectively.

Linear mixed effects (LME) statistical analysis via restricted maximum likelihood (REML) was applied to every condition on both the tracking error and cocontraction, in order to assess performance stability and evaluate whether the participants had adapted to noise. The model was fitted with the trial number as a fixed slope (\textit{s}) and a random intercept for each grouping factor (subject ID). {\color{black} For visual or haptic feedback experiment, the Shapiro-Wilk test showed that the cocontraction was normally distributed while the tracking error was not.} For visual and haptic feedback experiment, the Shapiro-Wilk test showed that neither the tracking error nor the cocontraction was normally distributed. {\color{black}For metrics with non-normal distribution, repeated measures ART ANOVA was used to analyse the effect of the visual and haptic noise}, the paired Wilcoxon signed-rank test was used for {\color{black}post-hoc} non-parametric hypothesis testing. For metrics with normal distribution, repeated measures ANOVA was used to assess the impact of visual or haptic noise. Paired T-test was conducted for post-hoc comparison between groups. The p-values for all comparisons were adjusted using the Holm-Bonferroni method.

\subsection{Visuo-haptic noise model}
The CNS perceives the target movement through visual feedback and haptic connection to the target, which is known to degrade the signal quality \cite{Takagi2018}. Assuming that this effect results in independent noise, the standard deviation of internal sensory noise is
\be \label{eq_HNModel}
\sigma^2_t(u) = \, \sigma^2_v \, + \, \sigma^2_{\kappa}(u) \,,
\ee
where $\sigma_v$ is the deviation of visual noise. Experiment data exhibit a saturation of tracking error with the increase of visual fuzziness (the size of the cloud) so the visual noise is modelled as
\begin{equation} \label{eq:regressionSig}
    \sigma_v = \alpha_v + \frac{\beta_v}{1+e^{-\sigma_c}}\,, \quad \alpha_v, \beta_v \in \mathcal{R}\,.
\end{equation}
where $\sigma_{c}$ is the angular deviation of target cloud given in the experiment. $\sigma_{\kappa}(u)$ is the deviation due to the joint compliance decreasing with muscle coactivity \cite{Takagi2018}, which was identified in \cite{carboni2021adapting} as
\be \label{eq_sigmac_exp}
\sigma_{\kappa}(u) = 5.18 \,+\, 49.65 \,e^{-6.11 u} \, .
\ee
Furthermore, the tracking error increases with the amplitude of haptic perturbation $\sigma_p$  thus a {\color{black} quadratic} regression model is used for the haptic noise: {\color{black}
\be \label{eq:regressionPoly}
\sigma_h = \,\alpha_p +\beta_p\sigma_p\,+\delta_p \sigma_p^2\,, \quad \alpha_p, \beta_p, \delta_p \in \mathcal{R}\,.
\ee}
The parameters of the computational model are identified using the cocontraction data of the last trial for all noise conditions. The effective visual noise deviation $\sigma_v$, the effective haptic noise $\sigma_h$ were identified by minimising the variation of the cost derivative to satisfy the first-order necessary optimal (Karush–Kuhn–Tucker) conditions \cite{boyd2004convex}. Considering the relationship between the deviation $\sigma_{\kappa}$ and the wrist's viscoelasticity, the visual noise deviation and the haptic noise deviation each has three values, resulting in six parameters to identify:
\be \label{eq_DiffCost}
\xi^* \equiv  \underset{\xi}{\arg\min} 
\bigg\{ \displaystyle \sum^3_{i=1} \sum^3_{j=1} \!\left[ \frac{\partial V}{\partial u} \left(u_{ij}(9), \, \sigma_{v}^{(i)},\, \sigma_h^{(j)}\right) \right]^2 \!\bigg\}
\nonumber
\ee
% \sigma_{v}^{o*}, \sigma_{v}^{w*}, \sigma_{v}^{s*}\!,\sigma_e^{o*}\!,\sigma_e^{w*}\!, \sigma_e^{s*}
% \sigma_{v}^o, \sigma_{v}^w, \sigma_{v}^s, \sigma_e^o, \sigma_e^w, \sigma_e^s

Using the collected cocontraction data $\{u_{ij}(9)\}$, a grid search is performed to determine the effects of visual and haptic noise under \{sharp, weak, strong\} conditions $\xi = \{\sigma_v^0, \sigma_v^w, \sigma_v^s, \sigma_e^0, \sigma_e^w, \sigma_e^s\}$ {\color{black} Particle swarm optimisation (PSO) \cite{kennedy1995particle} is employed within a bound range of [0,70],} yielding the noise values shown in Table \ref{Table:noise}. The optimal effort ratio {\color{black} $\gamma^*$=\,2.26} is then computed as the solution of
\be
0 \equiv \, \frac{d}{d\gamma} \!\left( \displaystyle \sum^3_{i=1}\sum^3_{j=1} \!\left[ \frac{\partial V}{\partial u} (u_{ij}(9), \sigma_{v}^{(i)}, \sigma_{e}^{(j)})\right]^2 \right).
\ee

A least-square regression using the identified parameters {\color{black} $\alpha_v$\,=\,-1.21,\, $\beta_v$\,=\,66.18} in eq.\,(\ref{eq:regressionSig}) was used to express the relationship between the angular deviation of the visual cloud $\sigma_{c}$ and effective deviation of visual noise $\sigma_{v}$. {\color{black} A quadratic regression with $\alpha_p$\,=\,5.05,\, $\beta_p$\,=\,6.84,\, $\delta_p$\,=\,41.68} was identified to model the relation between the perturbation amplitude and effective haptic noise.

\section*{Acknowledgment}
We thank Dagmar Sternad for discussions on the experimental protocol, and the participants for taking part in the experiment. 

% \section*{Author Contributions}
% The contributions are as follows:
% \begin{itemize}  
% \item Conceptualization:	
% XC, GC, AT, EB. 
% \item Data Curation: XC, SS. 
% \item Formal Analysis: XC, SS, EI.
% \item Funding Acquisition: EB. 
% \item Investigation: XC, SS.  
% \item Methodology: XC, SS, EI, GC, AT, EB.
% \item Project Administration: XC, EB.
% \item Software: XC, SS, EI.
% \item Supervision: EB.
% \item Visualisation: XC, SS, EI, AT, EB.
% \item Writing – Original Draft Preparation: XC.
% \item Writing – Review \& Editing: XC, SS, EI, AT, EB.
% \end{itemize}

% \section*{Competing interests}
% The authors have declared that no competing interests exist.

\bibliographystyle{IEEEtran}
\bibliography{IEEEabrv,Mendeley}

\end{document}